\documentclass[prb,aps,twocolumn,superscriptaddress,showpacs,floatfix]{revtex4-2}

\usepackage{graphicx}
\usepackage{epstopdf}
\usepackage[colorlinks=true,linkcolor=black, citecolor=blue, urlcolor=blue,
    unicode=true]{hyperref}
\usepackage{gensymb}
\usepackage{booktabs}
\usepackage{dcolumn}
\usepackage{bm}
\usepackage[dvipsnames]{xcolor}
\usepackage{float}
\usepackage{setspace}
\usepackage{comment}
\usepackage[section]{placeins}
\usepackage{siunitx}

\begin{document}

\preprint{APS/123-QED}

\title{Impurity-induced moment freezing in NaFe$_x$Ru$_{1-x}$O$_2$}

\author{Alon Hendler Avidor}
 \email{hendleravidor@ucsb.edu}
 \affiliation{Materials Department, University of California Santa Barbara, Santa Barbara, CA 93106, USA}

\author{Brenden R. Ortiz}
 \affiliation{Materials Department, University of California Santa Barbara, Santa Barbara, CA 93106, USA}
\affiliation{Materials Science and Technology Division, Oak Ridge National Laboratory, Oak Ridge, 37831, Tennessee, USA}

 \author{Paul M. Sarte}
 \affiliation{Materials Department, University of California Santa Barbara, Santa Barbara, CA 93106, USA}
 
  \author{Qiang Zhang}
 \affiliation{Neutron Scattering Division, Oak Ridge National Laboratory, Oak Ridge, Tennessee 37831, USA}
 
 \author{Stephen D. Wilson} 
 \email{stephendwilson@ucsb.edu}
 \affiliation{Materials Department, University of California Santa Barbara, Santa Barbara, CA 93106, USA}

\date{\today}

\begin{abstract}
We report the impact of magnetic impurity substitution on the quantum disordered magnetic ground state of NaRuO$_2$. Local $S=5/2$ moments are introduced into the frustrated triangular lattice of $J_{eff}=1/2$ Ru moments via Fe-substitution in NaFe$_x$Ru$_{1-x}$O$_2$, and the evolution of the magnetic ground state is reported.  Local spin freezing associated with conventional spin glass behavior is observed upon Fe substitution, marking an impurity-induced freezing of the primarily dynamic magnetic ground state in NaRuO$_2$.  Furthermore, local Fe moments induce a Curie-Weiss magnetic behavior in the uniform magnetic susceptibility, and the local moment magnitude is best described by dynamic Ru moments polarized about impurity sites.  Our results establish an impurity-doping phenomenology consistent with inherently dynamic moments in NaRuO$_2$ that are pinned by local magnetic impurities, similar to ``swiss cheese" models of impurity-substituted copper oxides, 
\end{abstract}

\maketitle

\section{Introduction}

  Layered oxides with the structure type $\alpha$-Na$M$O$_2$ where $M$ is a transition metal or lanthanide ion often host a frustrated triangular lattice of magnetic moments capable of stabilizing a variety of unconventional phenomena. The magnetic layers are composed of $M$O$_6$ edge-sharing octahedra, which are spaced by alkai metal layers.  This results in predominantly quasi-two dimensional interactions, and the local orbital and spin configurations of the $M$-site ion can drive a wide variety of collective magnetic behaviors.  For instance, in the limit of weak spin-orbit coupling, the relatively isotropic $S=1/2$ case of $M$=Ti is proposed to form a nonmagnetic orbital dimer state \cite{PhysRevLett.94.156402} whereas for $S=5/2$ moments and $M$=Fe competing exchange interactions drive a complex thermal evolution of magnetic order and short-range correlations \cite{PhysRevLett.94.156402}.  Orbital effects, such as a strong Jahn-Teller distortion in the $S=2$, $M$=Mn case, can drive the formation of quasi-one dimensional spin correlations and stabilize single-ion anisotropy-driven magnon bound states \cite{dally2020three, dally2018amplitude, stock2009one}, and orbital order can also intertwine with magnetic ordering in the case of $S=1$ and $M$=V \cite{NaVO2}.

  The addition of appreciable spin-orbit coupling can introduce further richness into the phase diagrams of these materials.  Intrinsically quantum disordered or quantum spin liquid ground states have been reported in a number of spin-orbit entangled $M$=lanthanide compounds with $S_{eff}=1/2$ ground state wave functions \cite{paddison2017continuous, shen2016evidence, bordelon2019field, ding2019gapless}.  For instance, a number of intrinsically quantum disordered magnetic ground states form in NaYb$X_2$ ($X$=O, S, Se) compounds \cite{bordelon2019field, baenitz2018naybs, dai2021spinon, ranjith2019anisotropic} and in RbCeO$_2$ \cite{ortiz2022electronic} with well-isolated Kramers doublet ground state wave functions \cite{zhang2021crystalline, bordelon2020spin, bastien2020long}. Moving toward the limit of stronger crystal electric fields, the spin-orbit Mott state in NaRuO$_2$ also realizes a spin-orbit entangled $J_{eff}=1/2$ ground state wave function built from $t_{2g}$ orbitals with strong anisotropic exchange interactions \cite{ortizsarte}.  Despite predictions of a ferromagnetic ground state \cite{razpopov2023j,bhattacharyya2023naruo2}, an unusual, quantum disordered magnetic ground state forms in this compound---a state marked by persistent low-frequency dynamics and a field-tunable $\gamma$-term in the low temperature specific heat.  
  
  While the correct starting Hamiltonian for understanding the formation of quantum disorder in NaRuO$_2$ remains a topic of investigation, with candidates ranging from Heisenberg-Kitaev-$\Gamma$ models \cite{trebst2022kitaev} to charge fluctuations inherent to a weak Mott state \cite{PhysRevLett.111.157203, PhysRevB.81.245121} to extended/ring-exchange effects across the triangular network \cite{motrunich2005variational}, the fate of local moments within this compound also remains an open question. High-temperature susceptibility data in NaRuO$_2$ is dominated by a weakly temperature-dependent Van Vleck response that does not fit to a Curie-Weiss formalism below 400 K \cite{ortizsarte}.  This is likely due to a combination of single-ion Kotani-type effects combined with a strong mean-field term, an analysis challenge that has only recently been addressed in strongly spin-orbit coupled transition metal halides \cite{li2024origins}. Nevertheless, a partial freezing of moments is observed below $\approx 2$ K in NaRuO$_2$, and the origin of this freezing remains an open question.

\begin{figure*}[]
    \centering
   \includegraphics[width=\textwidth]{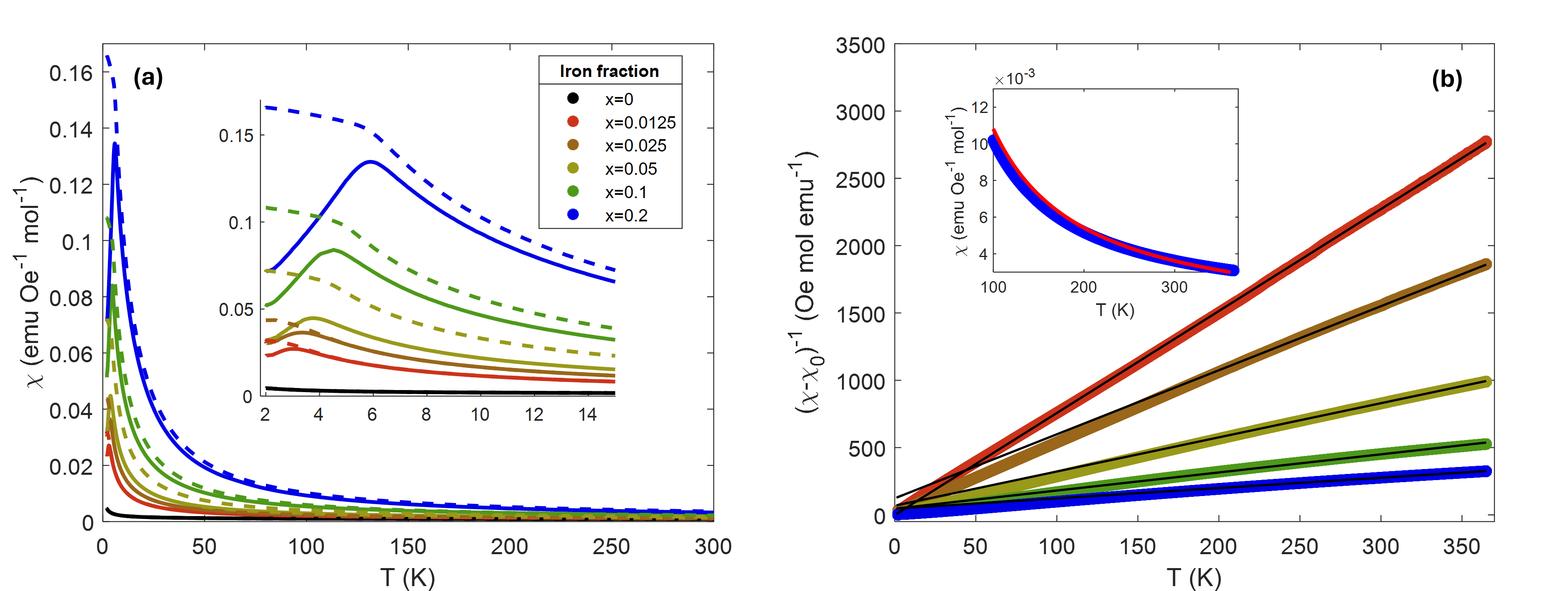}
    \caption{DC magnetic susceptibility, $\chi$, collected under H=200 Oe for the NaFe$_x$Ru$_{1-x}$O$_2$ series. (a) $\chi$ plotted as a function of temperature showing zero field cooled (ZFC, solid line) and field-cooled (FC, dashed line) data, as well as a peak for the ZFC data in all samples. The inset shows an expanded low temperature region, $2<T<15$ K for greater detail. (b) $\chi^{-1}$ as a function of temperature collected between 2 K and 365 K. The linear fits are to a Curie-Weiss form of susceptibility as described in the text. The inset shows the fit to the form $\chi=\chi_0+C/(T-\Theta_{CW})$ $\chi_0$ is a temperature independent term fit across the entire temperature range.}
    \label{fig:crystal structure}
\end{figure*}

  In particular, whether the weak freezing is driven by a small fraction of impurities or is an intrinsic property of the bulk system are yet to be determined.  Prior muon spin relaxation measurements suggest that the freezing occurs throughout the sample \cite{ortizsarte}; however nonmagnetic defect substitution in the form of local Na defects within the Ru sublattice of Na$_{3+x}$Ru$_{3-x}$O$_6$ also drive clear spin glass freezing with a clear signature in the irreversibility of the susceptibility \cite{ortiz2022defect}. Traversing the solid solution Na$_{3+x}$Ru$_{3-x}$O$_6$, however, is a strong perturbation to the system, where the oxidation state of Ru is altered in parallel to the introduction of in-plane defects. An alternative approach is to introduce magnetic, in-plane disorder into NaRuO$_2$ to test the potential for defect-induced spin freezing and also to test the influence of a local moment on the background spin fluctuations.
  
 To address this problem, in this paper we present a study of the magnetic behavior of NaRuO$_2$ as isovalent Fe$^{3+}$ local moment impurities are introduced into the Ru$^{3+}$ triangular lattice network.  NaRu$_{1-x}$Fe$_x$O$_2$ samples were created with Fe concentrations ranging between $0\leq x\leq 0.2$, and the evolution of their magnetic properties is explored. Similar to the case of Na-defects introduced into the Ru-planes, Fe-impurities induce a conventional spin freezing at low temperatures, whose onset temperature increases with the degree of disorder.  The $3d^5$ Fe$^{3+}$ $S=5/2$ local moment defects further induce a local moment magnetic response at high temperatures that can be modeled within a conventional Curie-Weiss formalism.  Surprisingly, the magnitude of the local moment exceeds that of isolated Fe$^{3+}$ moments and can instead be modeled assuming polarized neighboring Ru$^{3+}$ moments pinned about the impurity sites. Our results suggest that local moments are present in NaRuO$_2$ and are bound by either fluctuations or a strong exchange field.  Their liberation via in-plane impurities is reminiscent of ``swiss cheese" models of impurity-substituted cuprates \cite{nachumi1996muon}.

\section{Experimental details}
A series of polycrystalline NaFe$_x$Ru$_{1-x}$O$_2$ samples were synthesized using previously reported solid state techniques \cite{ortiz2022defect}. Starting mixtures were modified such that RuO$_2$ was replaced with Fe$_2$O$_3$ (Alfa Aesar, 99.99\% purity) and the ratio of Na$_2$O$_2$ and Na was adjusted to maintain stoichiometry for each target concentration $x$.  Five samples with $x$ = 0.0125, 0.025, 0.05, 0.1 and 0.2 were created, and attempts at larger concentrations of Fe showed signatures of phase separation and a potential miscibility limit of Fe within the matrix. To verify the chemical composition and crystal structure, x-ray fluorescence spectroscopy measurements were performed (Rigaku ZSX Primus IV) and powder x-ray diffraction measurements (Panalytical Empyrean, Cu $\lambda = 1.54 (\si{\angstrom}) $) were performed \cite{suppinfo}.  These measurements confirmed that the structure maintained a single phase $R\bar3m$ space group as a function of Fe composition and that the target substitution level $x$ was realized in the final powders. Attempts to prepare the sample series using an alternative method of the reported process for NaFeO$_2$ \cite{mcqueen} while exchanging a portion of reagent Fe$_2$O$_3$ with RuO$_2$ were unsuccessful. Both DC and AC susceptibility measurements were conducted using a Quantum Design Magnetic Properties Measurement System (MPMS). For AC-susceptibility measurements, all samples were measured with a driving field of 3 Oe at 10, 100, 250, 500 and 700 Hz. Neutron diffraction data were collected using the POWGEN diffractometer at the Spallation Neutron Source at Oak Ridge National lab. Neutron frames 2 and 3 with center wavelengths of 1.5 (\si{\angstrom}) and 2.665 (\si{\angstrom}) were used, respectively, for the data collection. An orange cryostat was adopted as the sample environment to cover the temperature region of 1.6-300 K.

\begin{figure}[]
	\centering
	\includegraphics[width=0.5\textwidth]{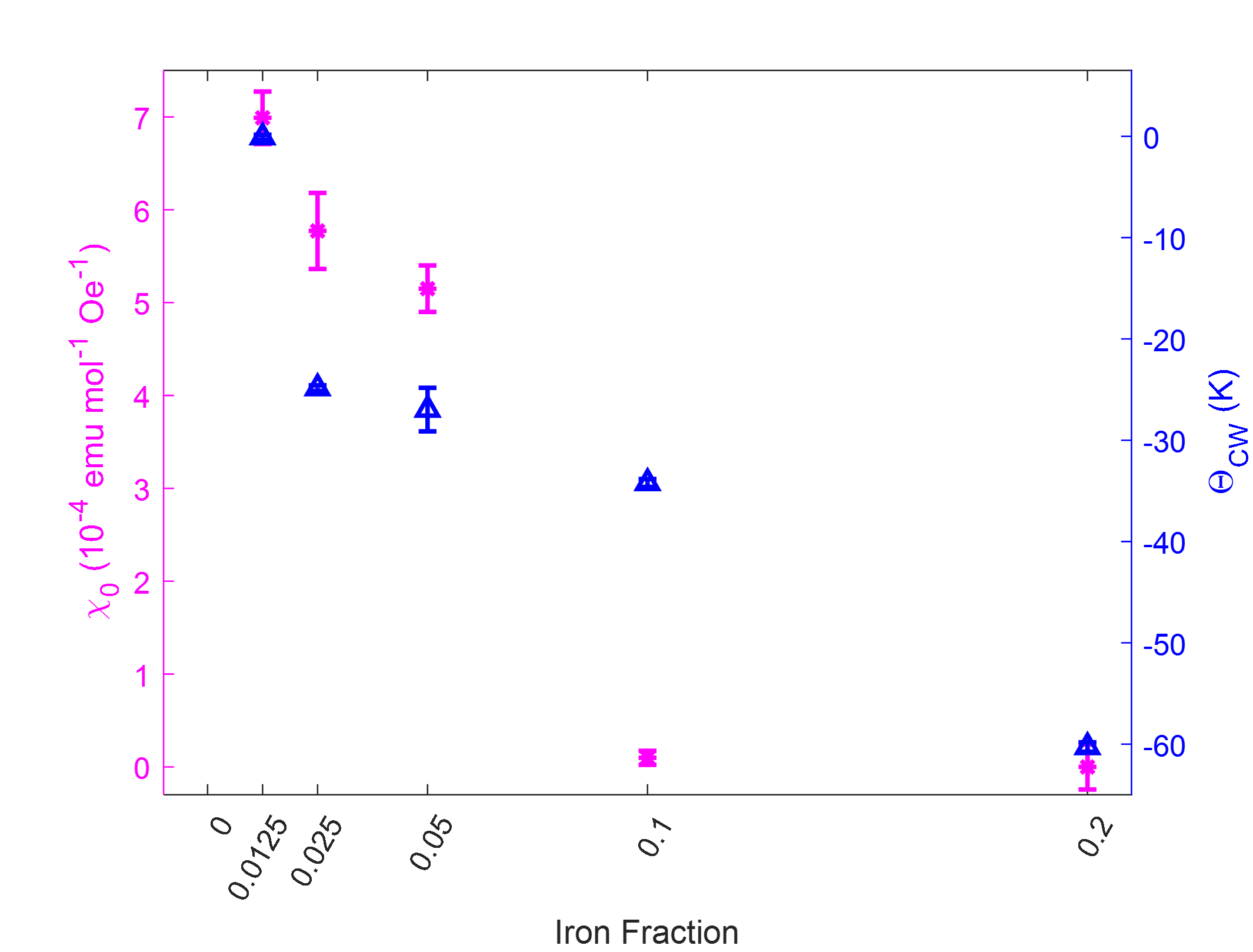}
	\caption{The Curie-Weiss temperatures $\Theta_{CW}$ and temperature independent susceptibility $\chi_0$ extracted from the fits shown in Figure 1 across the entire the NaFe$_x$Ru$_{1-x}$O$_2$ sample series.}
	\label{fig:Chi0andthetaCW}
\end{figure}

\vspace{2mm}
\begin{table} [h]
	\begin{ruledtabular}
		\begin{tabular}{lcccr}
			\textrm{$x$}&
			\textrm{$T_f$ $(K)$}&
			\textrm{$\mu_{eff}$ $(\mu_B)$}&
			\textrm{$\mu_{eff}$ $(\mu_B)$}&
			\textrm{$\mu_{eff}$ $(\mu_B)$}\\
			\textrm{Fe content}&
			\textrm{}&
			\textrm{measured}&
			\textrm{Fe only}&
			\textrm{Ru and Fe}\\
			\colrule
			0.0125  & 2.93& 1.03 $\pm$ 0.03& 0.66& 0.93\\
			0.025 & 3.32& 1.36 $\pm$ 0.05&0.93& 1.29\\
			0.05 & 3.80& 1.78 $\pm$ 0.06&1.32& 1.74 \\
			0.10 & 4.52& 2.43 $\pm$ 0.06&1.87& 2.34 \\
			0.20 & 5.93& 3.23 $\pm$ 0.02& 2.64& 3.05\\ 
			1 & 11& 5.80 Ref. \cite{ichida}& 5.91 & - \\
		\end{tabular}
	\end{ruledtabular}
	\caption{Spin freezing temperatures and effective magnetic moments measured for all NaFe$_x$Ru$_{1-x}$O$_2$ samples. Two models of a paramagnet are added for comparison--- the first with only the Fe $S=5/2$ spins, and the second with the Fe spins as well as induced Ru $J_{eff}=1/2$ moments. The model assumes account the first and second  nearest neighbors polarizing, as discussed in the text. For pure NaFeO$_2$, the values are taken from Ichida et al.\cite{ichida}}
\end{table}

\section{Experimental Results}

DC magnetization measurements were performed to study the evolution of the magnetic behavior in NaFe$_x$Ru$_{1-x}$O$_2$ with increasing $x$. Low-field susceptibility (M/H) data were collected upon warming from 2 K to 300 K at a sweep rate of 2 K/min, using a DC field of 200 Oe after both zero-field cooling (ZFC) and field cooling (FC) conditions.  The resulting data are shown in Figure 1 (a). Two qualitative changes are immediately  evident in the susceptibility with the introduction of Fe impurities.  The first is that magnetic susceptibility is significantly enhanced at all temperatures due to an added Curie-like response, and the second is the appearance of a peak in ZFC conditions which shows irreversibility between FC and ZFC curves. The low temperature behavior is consistent with a disorder-driven spin glass freezing, which is absent in pure NaRuO$_2$ over this temperature regime. For reference, $\alpha$-NaFeO$_2$ orders below $T_N=11.1$ K \cite{ichida}, and a subtle irreversibility is also reported at the same temperature, likely due to residual fluctuations into the lower temperature magnetic state \cite{mcqueen}.  

The impurity-induced, high-temperature magnetic susceptibility can be fit via a Curie-Weiss model with a temperature-independent $\chi_0$ term added. Figure 1 (b) shows the resulting $(\chi(T)-\chi_0)^{-1}$ data plotted with the fit $\chi_0$ term removed. A fit range between 200 K and 365 K was used for all samples and linear fits were possible for the entire NaFe$_x$Ru$_{1-x}$O$_2$ series. This reflects the introduction of local moments via the addition of Fe impurities. The effective magnetic moment extracted from the Curie-Weiss fits can then be compared with the expectation of introducing a fraction $x$ of $S=5/2$ moments, while the host material NaRuO$_2$ does not show clear Curie-Weiss behavior \cite{ortizsarte}.  Figure 1 (a) also shows the susceptibility for the undoped $x=0$ compound for comparison.


As summarized in Table 1, the simplest model of introducing isolated $S=5/2$ local moments fails to explain the magnitude of Curie-Weiss moment. The local moment $\mu_{eff}=g\sqrt{NJ(J+1)}$ in units of $\mu_B$, where N is the volume fraction of the $x$ magnetic moments assuming the previously reported $g=2$ \cite{ichida} and $J=5/2$, is systematically larger throughout the NaFe$_x$Ru$_{1-x}$O$_2$ series.  One possible origin of this response is to assume that local moments are present in NaRuO$_2$, yet bound by strong interactions or hidden by strong fluctuations in the measurable temperature regime.  The introduction of spin impurities into the Ru triangular lattice can locally freeze and polarize the $J_{eff}=1/2$ Ru$^{3+}$ moments and generate an enhanced $\mu_{eff}$ per magnetic impurity added.

To test this possibility, a simple Monte Carlo model of Fe$^{3+}$ spins spread randomly across the triangular lattice was created, assuming neighboring Ru$^{3+}$ moments are polarized out to the nearest neighbor and next-nearest neighbor distances. In this model, assuming each Ru $J_{eff}=1/2$ moment contributes $\mu_{eff}=1.73$ $\mu_{B}$, the best fit to the total effective moment extracted from Curie-Weiss fits results when Ru moments are polarized up to the next-nearest neighbor about each Fe impurity site. This Monte Carlo approach  allows for an approximation of the substitution level at which all Ru moments are polarized at various coordination levels about the Fe-sites and for the determination of the relative fraction of Ru moments polarized at intermediate impurity substitution levels \cite{suppinfo}.

 \begin{figure*}[]
	\centering
	\includegraphics[width=\textwidth]{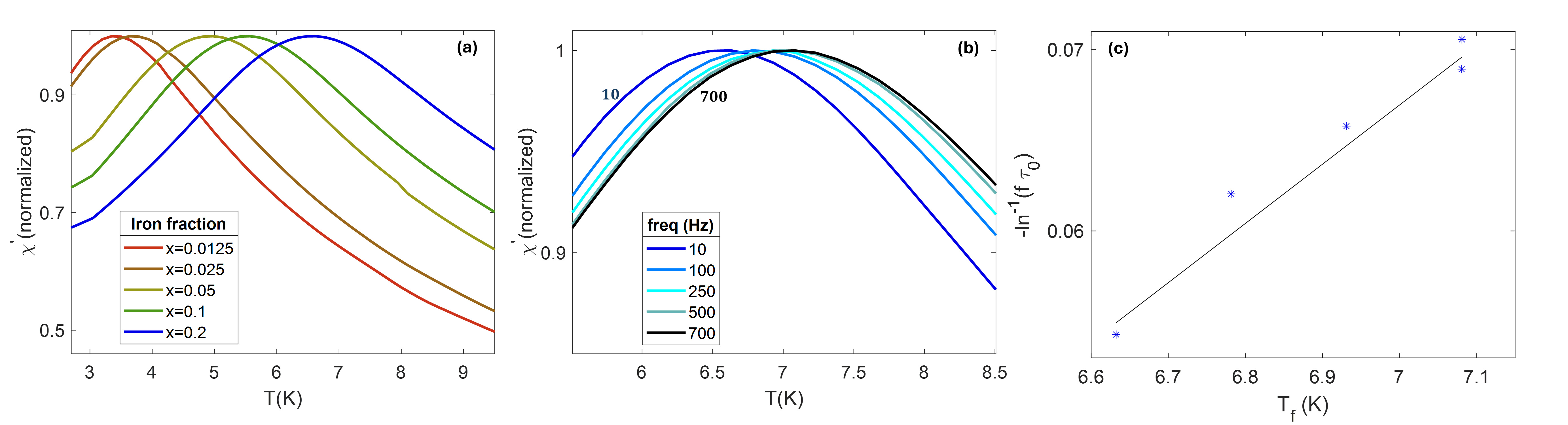}
	\caption{(a) The AC susceptibility of all samples collected at 10 Hz. The real part of the susceptibility $\chi'$ is normalized to 1 at its peak and plotted as a function of temperature. (b) The frequency dependence of the peak in $\chi^{\prime}$, labeled as $T_f$, is shown for the sample with the largest fraction of iron content, NaFe$_{0.2}$Ru$_{0.8}$O$_2$. $T_f$ is shifted upward in temperature upon increasing frequency. (c) The Vogel-Fulcher fit to the frequency dependence of $T_f$, shown for the NaFe$_{0.2}$Ru$_{0.8}$O$_2$ sample.}
	\label{fig:AC}
\end{figure*}

Figure 2 plots the evolution of the $\chi_0$ term and $\Theta_{CW}$ as a function of Fe-content $x$. $\Theta_{CW}$ shows a finite antiferromagnetic value that grows with increasing Fe content up to the $x=0.2$ limit.  Notably, at this limit, $\Theta_{CW}$ becomes substantial, reflecting strong interactions between the impurity-induced moment clusters.  As this large mean field appears, the  $\chi_0$ term simultaneously vanishes with increasing $x$.  The large $\chi_0$ of the parent $x=0$ compound \cite{ortizsarte} is rapidly quenched and vanishes by $x=0.1$, which is a concentration coinciding with nearly all Ru moments being polarized by local Fe impurities in the value for $\mu_{eff}$.  This is again consistent with a two-phase model of local Fe impurities polarizing an extended halo of Ru moments that rapidly percolates with $x$.


Now turning to the lower temperature regime of the measured susceptibility, the freezing behavior shown in the inset of Figure 1 (a) evolves as a function of $x$, and the onset of irreversibility $T_f$ is summarized in Table 1. This transition is better illustrated through AC susceptibility measurements plotted in Figure 3 (a) where the 10 Hz data are normalized to a common peak value and  $T_f$ is seen to reach a maximum of $\approx6$ K at $x=0.2$.  To illustrate the spin freezing further, the frequency dependence of $\chi^{\prime}(T)$ for the $x=0.2$ composition is plotted in Figure 3 (b).  A clear shift in the cusp is seen upon increasing frequency, which can be parameterized via a standard Vogel-Fulcher analysis \cite{shtrikman,AHARONI}. Fits were performed to the form: \begin{equation}
	\tau= \tau_0e^{(E_a/{k_B(T_f-T_0)}}
\end{equation}
Here $f=\tau^{-1}$ is the frequency, $T_f$ is the relaxation (freezing) temperature, and there are three phenomenological, system-dependent parameters: characteristic relaxation time $\tau_0$, the activation energy $E_a$, and the Vogel-Fulcher temperature $T_0$, a measure of intercluster interaction strength \cite{anand}. 

 Figure 3(c) shows the result of fitting data in panel 3(b) to the Vogel-Fulcher relation. The activation energy can be deduced from the slope in Figure 3 (c): $E_a/k_B=30.6$ K which is consistent with reported values for other Fe-based spin glasses \cite{Benka,chandra,vijaya}. $\tau=10^{-9}$ was obtained, a value more in line with cluster spin glass systems, and larger than atomic spin glass \cite{khurshid,Mukherjee,chandra,vijaya}.  

To explore the most heavily doped compound, $x=0.2$, further, neutron powder diffraction measurements were performed.  Data were collected across a series of temperatures down to 1.6 K, and no magnetic or crystallographic transitions were observed. The $E_i=1.5$ \si{\angstrom} measurements for all temperatures are shown in Figure 4.  Consistent with a spin glass state below $T_f$, no magnetic superlattice reflections were observed or correlated magnetic intensity evident below 6 K. Magnetic diffuse scattering was also not resolved, though the small fraction of $S=5/2$ moments would likely preclude its observation. Although no long-range magnetic order was observed, crystallographic information can be determined from the diffraction data, and the lattice and unit cell parameters are summarized in Table II.

\begin{figure}[b]
    \centering
    \includegraphics[width=0.5\textwidth]{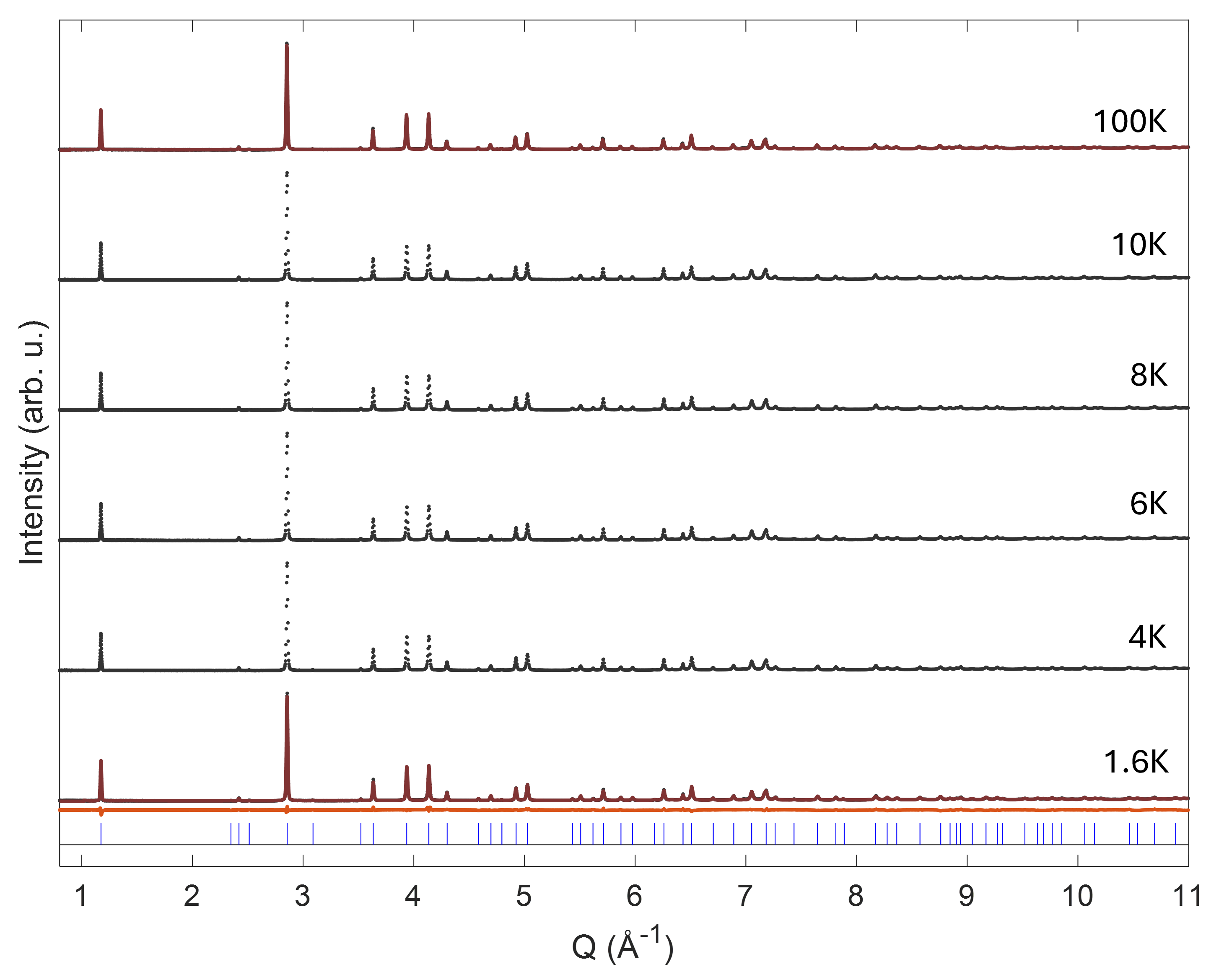}
    \caption{Neutron diffraction dataq collected at different temperatures for the iron rich sample NaFe$_{0.2}$Ru$_{0.8}$O$_2$. Refinement results of the structure are shown for $T=1.6$ K and $T=100$ K data (red), and the difference between the data and the refined structural model is shown for $T=1.6$ K below the data. Tick marks indicate the expected Bragg peak positions for the NaRuO$_2$ phase.}
    \label{fig:neutrondiffraction}
\end{figure}

\begin{table} []
\begin{ruledtabular}
\begin{tabular}{lccr}
\textrm{}&
\textrm{NaRuO$_2$}&
\textrm{NaFe$_{0.2}$Ru$_{0.8}$O$_2$}&
\textrm{NaFeO$_2$}\\
\colrule
$T$ (K) & 1.5 \cite{ortizsarte} & 1.6& 4.2 \cite{tomkowicz}\\ 
$a$ (\si{\angstrom}) & 3.055& 3.036&3.018\\
$c$ (\si{\angstrom}) & 16.122& 16.052&16.047 \\
\end{tabular}
\end{ruledtabular}
\vspace{2mm}
\begin{ruledtabular}
	\begin{tabular}{lccccr}
		\textrm{Atom}&
		\textrm{Wyckoff}&
		\textrm{x}&
		\textrm{y}&
		\textrm{z}&
		\textrm{Occupancy}\\
		\colrule
		Fe & 3b & 0 & 0 & 0.5 & 0.192\\ 
		Ru & 3b & 0 &0 &0.5 & 0.790\\
		Na  & 3a & 0 &0 & 0 & 0.925 \\
		O  & 6c & 0 & 0 & 0.23344(5)& 1.0000 \\
	\end{tabular}
\end{ruledtabular}

\caption{\label{latpar}The measured lattice parameters of NaFe$_{0.2}$Ru$_{0.8}$O$_2$, compared to its two stoichiometric end points: NaRuO$_2$ taken from Ortiz et al. \cite{ortizsarte} and NaFeO$_2$ from Tomkowicz et al. \cite{tomkowicz}. The temperatures with which the respective measurements were taken are also shown for comparison.  Unit cell parameters from Rietveld refinement of NaFe$_{0.2}$Ru$_{0.8}$O$_2$ with R$_{exp}=1.227$ and R$_{wp}=8.948$}
\end{table}

\section{Discussion and Conclusions}

The structure of NaFe$_{0.2}$Ru$_{0.8}$O$_2$ remains in the $R\bar3 m$ space group, the same as either side of the alloy series NaRuO$_2$ and NaFeO$_2$, both shown in Table II for comparison.  The magnetic states at either end of the alloy series are both frustrated by the triangular lattice motif of the transition metal ions; however, the more quantum $J_{eff}=1/2$ moments of NaRuO$_2$ manifest an intrinsically disordered ground state while the more classical moments of NaFeO$_2$ show a lock-in like transition from incommensurate to commensurate order upon cooling below 8 K \cite{mcqueen}.  The spin glass freezing of moments in the Fe substituted NaRu$_{1-x}$Fe$_x$O$_2$ is consistent with expectations of a disorder-driven freezing of the otherwise dynamic moments in NaRuO$_2$---an effect seen in a number of impurity-substituted spin liquid candidate materials \cite{ortiz2022defect, PhysRevMaterials.4.064410}. We note that this effect allows us to distinguish Fe incoporation into the lattice from trivial free Fe in the sample.  While there may be a small fraction of Fe moments unincorporated, it is well below the background of the measurement.  More extensive inclusions of Fe can be excluded via our diffraction measurements and the absence of detectable ferromagnetism at high temperature in isothermal magnetization measurements.

The susceptibility of NaRu$_{1-x}$Fe$_x$O$_2$ evolves in a manner suggestive of ``swiss cheese" models proposed for impurity-substituted high-temperature superconducting cuprates \cite{wakimoto2005magnetic, nachumi1996muon}.  In cuprates, nonmagnetic Zn substitution into the copper oxide planes drives a local suppression of the superconducting ground state into patches of normal-state metals with a Zn-induced polarization of local Cu moments. In NaRuO$_2$, the addition of magnetic Fe impurities into the Ru-plane seemingly locally suppresses the fluctuations of nearby Ru moments and polarizes them. 
 
 Monte Carlo simulations show that the susceptibility data are best fit with Ru moments polarized out to the next-nearest neighbors from the Fe-sites, leading to a relatively rapid polarization of all Ru moments by $\approx 10\%$ site-substitution.  This is consistent with the complete suppression of the $\chi_0$ term in susceptibility at $x=0.1$, and implies a connection between the enhanced $\chi_0$ value of the parent NaRuO$_2$ compound and its bound/fluctuating Ru-moments. In parallel to the suppression of the $\chi_0$ term, an enhanced, antiferromagnetic mean field $\Theta_{CW}$ is resolved.  This exchange field grows continuously in strength with added Fe content, consistent with a model of phase separation and percolating clusters of Fe-polarized Ru moments. Some caution should be given to the interpretation of the large $\Theta_{CW}$ values obtained for larger $x$ values.  These values begin approaching the start of the temperature regime used to perform Curie-Weiss fits to the data, and, in the extreme $x=1$ NaFeO$_2$ limit, there is debate whether a Curie-Weiss model can be applied \cite{mcqueen, tomkowicz}.  An additional caveat is that our analysis is based on an assumed full polarization of each neighboring Ru moment, whereas our measurements cannot distinguish between this scenario and a partial polarization of a larger halo of Ru moments about each Fe site.

The Ru-induced moment in Fe-substituted NaRuO$_2$ invites comparison to the well-studied impurity effects in cuprates \cite{RevModPhys.81.45}.  In the 2D copper oxide planes of cuprates, nonmagnetic substitution induces a local moment and the formation of a staggered polarization about the impurity site.  
A similar staggered polarization of some variety may form in NaRu$_{1-x}$Fe$_x$O$_2$ below its freezing temperature; however correlations above this temperature are not accessible with our current data.  A notable  difference between the ruthenate and cuprate cases is the nature of their parent Mott states, where within NaRuO$_2$, a ferromagnetic ground state is predicted \cite{razpopov2023j,bhattacharyya2023naruo2}.  This contrasts the antiferromagnetic Mott phase endemic to the cuprates and the nature of their impurity-induced spin correlations may differ. Another difference is the nature of the impurity dopants, where the magnetic impurity Fe differs from the nonmagnetic impurity-induced moments commonly studied in the cuprates.  The apparent reduction of the local moment in magnetic impurity studies of the cuprates \cite{PMendels_1999} versus the enhancement in NaRuO$_2$ likely arises from the differing local orbital physics of the $e_g$ electrons in cuprates versus the $t_{2g}$ orbital mixture in the ruthenate.

 In summary, we have synthesized a series of magnetically doped NaFe$_x$Ru$_{1-x}$O$_2$ compounds. All compounds maintain the crystallographic structure of the parent compound, and an immediate spin freezing appears at the lowest level of Fe substitution attempted ($x=0.0125$). Susceptibility measurements identify a spin glass state that emerges with Fe-doping with a freezing temperature that conventionally increases with increased impurity content. A larger local moment is observed per Fe$^{3+}$ ion than that expected for $S=5/2$ impurities, and instead the local moment is best modeled via Fe-impurities polarizing a radius of local $J_{eff}=1/2$ Ru-moments up to the next-nearest neighbor level.  Our findings suggest a model of magnetic impurities locally suppressing the fluctuating or dimer-bound magnetic ground state of NaRuO$_2$.  Our data further support the hypothesis that the partial freezing observed in pristine NaRuO$_2$ arises from residual amounts of disorder in powders, likely in the form of Na substitutional disorder within the RuO$_2$ planes.
\section*{Data Availability}
{The raw supporting the findings of this study are openly available in Zenodo with DOI:10.5281/zenodo.13750994}.

 \section{Acknowledgments}
  This work was supported by the US Department of Energy (DOE), Office of Basic Energy Sciences, Division of Materials Sciences and Engineering under Grant No. DE-SC0017752.  The research made use of the shared facilities of the NSF Materials Research Science and Engineering Center at UC Santa Barbara (DMR- 1720256). The UC Santa Barbara MRSEC is a member of the Materials Research Facilities Network. (www.mrfn.org). This work also used facilities supported via the UC Santa Barbara NSF Quantum Foundry funded via the Q-AMASE-i program under award DMR-1906325. 

\bibliography{bibliography}

\end{document}